\pdfoutput=1
\documentclass[unsortedaddress,prb,twocolumn]{revtex4}
\usepackage[pdftex]{graphicx}

\usepackage{amsmath}
\usepackage{bm}
\usepackage{amssymb}

\usepackage[bookmarks]{hyperref}

\newcommand{\beq}{\begin{equation}}
\newcommand{\eeq}{\end{equation}}
\newcommand{\be}[1]{\begin{equation}\label{#1}}
\newcommand{\ee}{\end{equation}}
\newcommand{\continue}{\nonumber \\ }
\newcommand{\bea}{\begin{eqnarray}}

\newcommand{\eea}{\end{eqnarray}}

\newcommand{\tr}{{\rm tr}\, }

\newcommand{\VectorIII}[3]{
\begin{pmatrix} {#1} \\
                {#2} \\
                {#3}    
\end{pmatrix}
}

\newcommand{\refeq}  [1] {Eq.~(\ref{#1})}

\newcommand{\refFig} [1] {Fig.~\ref{#1}}

\newcommand{\refSect}[1] {Sec.~\ref{#1}}

\begin{document}

\title{ Strain distributions in lattice-mismatched semiconductor core-shell nanowires   }

\author{Niels S\o ndergaard$^{1}$}
\author{ Yuhui He$^{2}$}
\author{Chun Fan$^{3}$}
\author{Ruqi Han$^{2}$}
\author{Thomas Guhr$^{1,4}$}
\author{H.~Q. Xu$^{5}$\footnote{electronic mail: hongqi.xu@ftf.lth.se}}

\affiliation{$^{1}$Division of Mathematical Physics, LTH, Lund University, S-22100 Lund, Sweden}

\affiliation{$^{2}$Institute of Microelectronics, Peking University, Beijing 100871, China} 

\affiliation{$^{3}$Computer Center of Peking University, Beijing 100871, China}
\affiliation{$^{4}$Fachbereich Physik, Universit\"{a}t Duisburg-Essen, 47048 Duisburg, Germany}
\affiliation{$^{5}$Division of Solid State Physics, Lund University, Box 118, S-22100 Lund, Sweden}

%\begin{flushright}
%{\large \tt \# 496}
%\end{flushright}

\maketitle

\section*{Abstract}
The authors study the elastic deformation field in lattice-mismatched
core-shell nanowires with single and multiple shells.  We consider
infinite wires with a hexagonal cross section under the assumption of
translational symmetry.  The strain distributions are found by
minimizing the elastic energy per unit cell using the finite element
method. We find that the trace of the strain is discontinuous with a
simple, almost piecewise variation between core and shell, whereas the
individual components of the strain can exhibit complex variations.

\section{Introduction}
\label{sec:Intro}
Nanowires have many applications. For example, nanowire-based biological sensors\cite{ZPCWL05}, chemical detectors\cite{WLCWHLL04}, solar
cells\cite{LGJSY05,ZWM07,SDWA07,WCZZYPM08}, LEDs\cite{HDL05}, field
emitters\cite{CDXWWYYWG02}, transistors\cite{GHFY06}, and electronic
logic gates\cite{HDCLKL01} have been demonstrated.  
To further tailor the properties of nanowires,
experimentalists have grown nanowire axial heterostructures, such as nanowire quantum
dots\cite{BOSPTMDWS02}, and radial heterostructures, such as core-shell
\cite{LGDWL02, LXTWL05,SKLPSTS05} and multi-shell\cite{MMF06} nanowires.
In particular, a difference in lattice constant in heterostructures
leads to intrinsic strain, which for nanowires can be incorporated into the material much easier than for thin
films due to the more effective strain relaxation at free
surfaces\cite{G06}.  This gives the opportunity for strain-engineering of
the electronic and optical properties\cite{GSB94,GSB95}.  The
intrinsic elastic deformation field in heterostructured nanowires is, therefore, an important and 
necessary input for further investigations of the properties of these heterostructured nanowires.

In this article, we discuss and calculate the strain field in lattice-mismatched
core-shell and multi-shell nanowires. We find rich behaviour of the
individual strain components, whereas their combination, the
trace of the strain tensor (the volumetric strain), shows much less
variation.  Our article is organized as follows. The model and theory
is introduced in \refSect{sec:Theory}. The results and discussion are
presented in \refSect{sec:Results}. Finally \refSect{sec:Summary}
contains the summary and conclusions.

\section{Theory and method}
\label{sec:Theory}

\begin{figure}[ht]
\includegraphics[bb = 0 0 484 443,width=6cm]{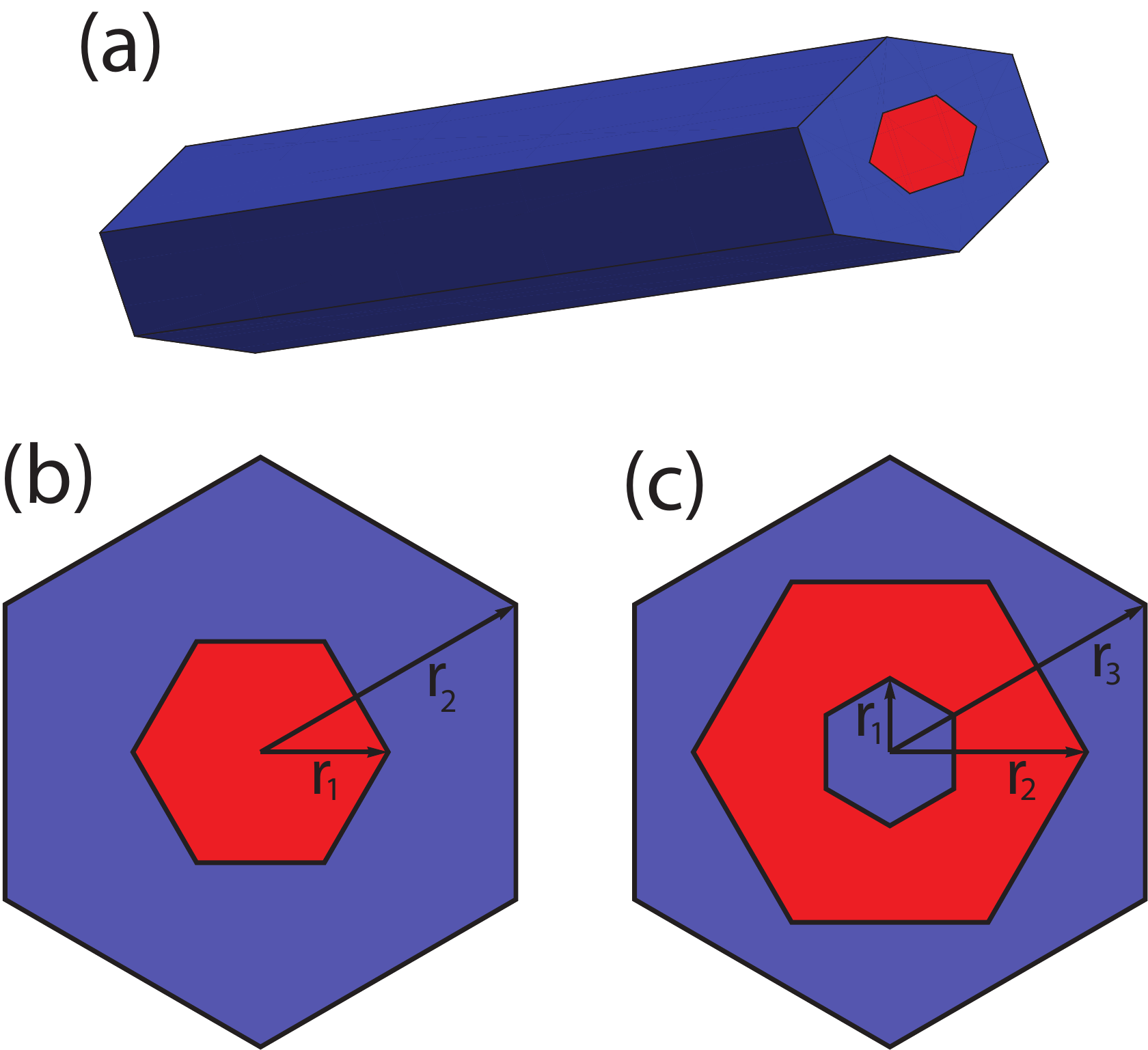}
\vspace{0cm}
\caption{(a) Schematic of a nanowire with a single shell around a core. (b) Cross section of the core-shell nanowire with core radius $r_1$ and shell radius $r_2$. (c) Cross section of a double-shell nanowire with core radius $r_1$ and shell radii $r_2$ and $r_3$.}
\label{fig:fig1}
\end{figure}

In this work, we calculate the elastic deformation field in  lattice-mismatched
core-shell and double-shell nanowires (see \refFig{fig:fig1} for schematics).  As an approximation, we
take the nanowires as infinite with translational symmetry along the growth direction.  
We argue below (based on Saint-Venant's principle\cite{Cl03}), that our
description of the infinite wires is also relevant for long finite 
wires. For simplicity, we neglect {\it exterior} forces acting on
surface or bulk parts of the wires, i.e., we consider  {\it free}
nanowires.  We restrict ourselves to nanowires with cubic lattice structure,  
but remark that generalizations to W\"urzite or other structures are straightforward.  
Furthermore, we consider only wires grown in the
$\left[ 111 \right]$-direction in this work. For convenience, we define a primed coordinate system
with respect to the growth direction, having basis vectors, 
\beq
\hat{\mathbf{x}}'=\frac{1}{\sqrt{2}} \VectorIII{1}{-1}{0},\hspace{1mm}
\hat{\mathbf{y}}'= \frac{1}{\sqrt{6}}\VectorIII{1}{1}{-2},
\hspace{1mm} \hat{\mathbf{z}}'=\frac{1}{\sqrt{3}}\VectorIII{1}{1}{1}
\,.  \eeq

Consider, for example, a core-shell nanowire with undeformed core and shell axial lengths of $L_c$
and $L_s$ and lattice constants of $a^{(c)}$ and $a^{(s)}$. To allow for
pseudomorphic matching, we shall assume that both core and shell have
the same number $N$ of unit cells in the axial direction and, thus, $L_c$ must
necessarily differ from $L_s$. We can write 
\beq
\label{eq:LayerCond}
\frac{L_c}{a^{(c)}} = \frac{L_s}{a^{(s)}} = C \cdot N \, , 
\eeq
where $C$ is some constant of proportionality.  In
the zincblende structure case  for  the $\left[111\right]$-direction, $C$ would be $\sqrt{3}$.

To  match the lattice of the shell to the core we introduce the  pseudomorphic initial
strain field $\boldsymbol{\varepsilon}^{(0)}$ in the shell. This choice of the pseudomorphic strain field initially scales all shell lattice vectors to have the same
length as in the core\cite{CT93}, 
\beq (1+\varepsilon^{(0)}) a^{(s)}
\equiv a^{(c)} \,,
\eeq 
and
\beq
\boldsymbol{\varepsilon}^{(0)} = \varepsilon^{(0)} \cdot \mathbf{1} \,
\, .
\eeq 
The total strain tensor is given by \cite{PdC06}
\beq
\boldsymbol{\epsilon} =\frac{1}{2}\left( \boldsymbol{\nabla}\otimes \mathbf{u}+ (\boldsymbol{\nabla}\otimes \mathbf{u})^t \right)+\boldsymbol{\varepsilon}^{(0)}  \,.
\eeq

For long beams, an often used approximation is that of {\it plane
strain}\cite{ZT94,Cl03}. In this approximation, the only non-zero strains are
$\epsilon_{x x}'$, $\epsilon_{yy}'$ and $\epsilon_{xy}'$, whereas
$\epsilon_{z x}'= \epsilon_{z y}'=0$ (planar sections remain flat) and $\epsilon_{z z}'=0$
(no axial extension). The theory has been  generalized to stretched beams by assuming that
the axial extension $\epsilon_{z z}'$ is non-vanishing but can be 
taken as a given fixed parameter\cite{L94}. 
For the core-shell and the double-shell nanowires we consider in this work,  $\epsilon_{z z}'$ is not known {\it a priori}, but we expect it to be present due
to the lattice mismatch. 
Therefore, we introduce an
$\epsilon_{z z}'$ for each sub-domain of a nanowire  and
consider these strains as variables. 
In the simple case of a core-shell structure with two sub-domains (the core and the shell), 
 the matching effectively reduces the two axial strains $\epsilon_{z z}'^{(c)}$
and $\epsilon_{z z}'^{(s)}$ to a single variable $a$,
\beq e_3'^{(i)}
\equiv \epsilon_{z z}'^{(i)} = \frac{a}{a^{(i)}}-1, \;\;\; \text{with}\; i=s,c \, . 
\eeq

In the material coordinate system,  the potential elastic energy is
\beq
\label{eq:StrainEnergy}
U = \int w \, dV \equiv \frac{1}{2} \, \int \mathbf{e} \cdot \mathbf{C} \cdot \mathbf{e} \, dV  \,,
\eeq
where $w$ is the strain energy density, $\mathbf{C}=\left[C_{i j}\right]$ is  the matrix of elastic constants and $\mathbf{e}=\left[ e_i \right]$ are engineering strains \cite{AM76}. 
In the nanowires with cubic lattice structure, the elastic constants are given by  three independent constants taken by convention as $C_{11},C_{12} \text{ and } C_{44}$. 

From now on, we shall only work in the transformed coordinate system and drop all primes in the notation.
In this coordinate system, we find, after some calculation, the energy density in \refeq{eq:StrainEnergy} as
\bea
\label{eq:UnstrainedEnergy}
w &=& \frac{1}{2} \left( D_1 (e_1^2 + e_2^2) + D_2 e_3^2 + D_3 e_1 e_2 \right.  \\ \nonumber & &+  \left. D_4 (e_1 + e_2) e_3 + D_5 e_6^2 \right) \, ,
\eea
with the constants
\bea
D_1 &=& \frac{1}{2} \left(C_{11}+C_{12}+2 C_{44}\right) \,,\\ \nonumber
D_2 &=& \frac{1}{3}  \left(C_{11}+2 C_{12}+4 C_{44}\right) \, , \\ \nonumber
D_3 &=& \frac{1}{3}  \left(C_{11}+5 C_{12}-2 C_{44}\right)\,, \\ \nonumber
D_4 &=&   \frac{2}{3}   \left(C_{11}+2 C_{12}-2 C_{44}\right)\,, \\ \nonumber
D_5 &=&   \frac{1}{6}   \left(C_{11}-C_{12}+4 C_{44}\right) \,.
\eea
The potential energy allows us to find the deformation field by a variational principle.  Minimizing the energy in \refeq{eq:StrainEnergy} leads to the equations
for a  collection of springs with generalized loads derived from the matching strain\cite{ZT94}.

An important ingredient of this work is the introduction of the axial extension as a variable. We shall therefore briefly discuss the variational procedure for that degree of freedom. Now consider a core-shell wire. Axial variations obey
\beq
\label{eq:VariationR}
\delta \epsilon_{z z}^{(i)} = \frac{\delta a}{a^{(i)}},  \;\;\; \text{with} \; i=s,c  \,.
\eeq
Varying the energy per unit cell with respect to the axial parameter $a$ gives
\bea
\label{eq:AxialVariation}
0 &=& \frac{1}{N}\,\frac{\delta U}{\delta a} \continue &=& \frac{1}{N}\,\left( \frac{1}{a^{(c)}} \int_c
\sigma_{zz} dV +\frac{1}{a^{(s)}} \int_s \sigma_{zz} dV \right) \continue &=&
\frac{1}{N}\,\left(
\frac{L_c}{a^{(c)}} \int_c \sigma_{zz} dS + \frac{L_s}{a^{(s)}} \int_s
\sigma_{zz} dS \right) \continue &\propto& \int_c \sigma_{zz} dS + \int_s
\sigma_{zz} dS \continue &=& \int_{c+s} dF_z = F_z \, ,\eea 
where we recognize that  the axial stress is given in the form of  
\beq 
\label{eq:SigmaZZ}
\sigma_{zz} \equiv \frac{\partial
w}{\partial e_{3}} = D_2 e_3 + \frac{D_4}{2} (e_1+e_2) \,.
\eeq 
Here we have used the assumption of translational invariance in the
$z$-direction. Also, we note that in the second equality of Eq.~(11), the integration in the axial direction
has been taken over the undeformed domains of the shell
and core. In \refeq{eq:AxialVariation},  $\sigma_{z z} dS$  is the vertical force $dF_z$ on an area
element $dS$. Hence, physically, \refeq{eq:AxialVariation} 
implies  the vanishing of  the total force in the axial
direction.    In the
multi-shell nanowire case, the condition \refeq{eq:AxialVariation} generalizes
to
\beq
\label{eq:CondAxialGen}
0=\sum_i \int_{\Omega_i} \sigma_{zz} \,dS \,.
\eeq

We remark that \refeq{eq:CondAxialGen}
should also apply to finite wires.  Thus, using the principle of
Saint-Venant\cite{Cl03}, we expect our results to describe well the strain
field in the middle sections of finite, but long, free wires.  If a net
total force and moment on the terminating ends are present, the
corresponding condition \refeq{eq:CondAxialGen} should be changed
accordingly.

In the numerics, we have used the linear finite element method\cite{ZT94}  to  
 minimize  \refeq{eq:StrainEnergy} with \refeq{eq:AxialVariation} imposed 
 for a free two-dimensional boundary.  The  planar degrees of freedom 
are seen to couple to the axial degree of freedom via, e.g., \refeq{eq:AxialVariation}.

\begin{figure}[ht]
\includegraphics[bb=0 200 500 748,width=8.5cm]{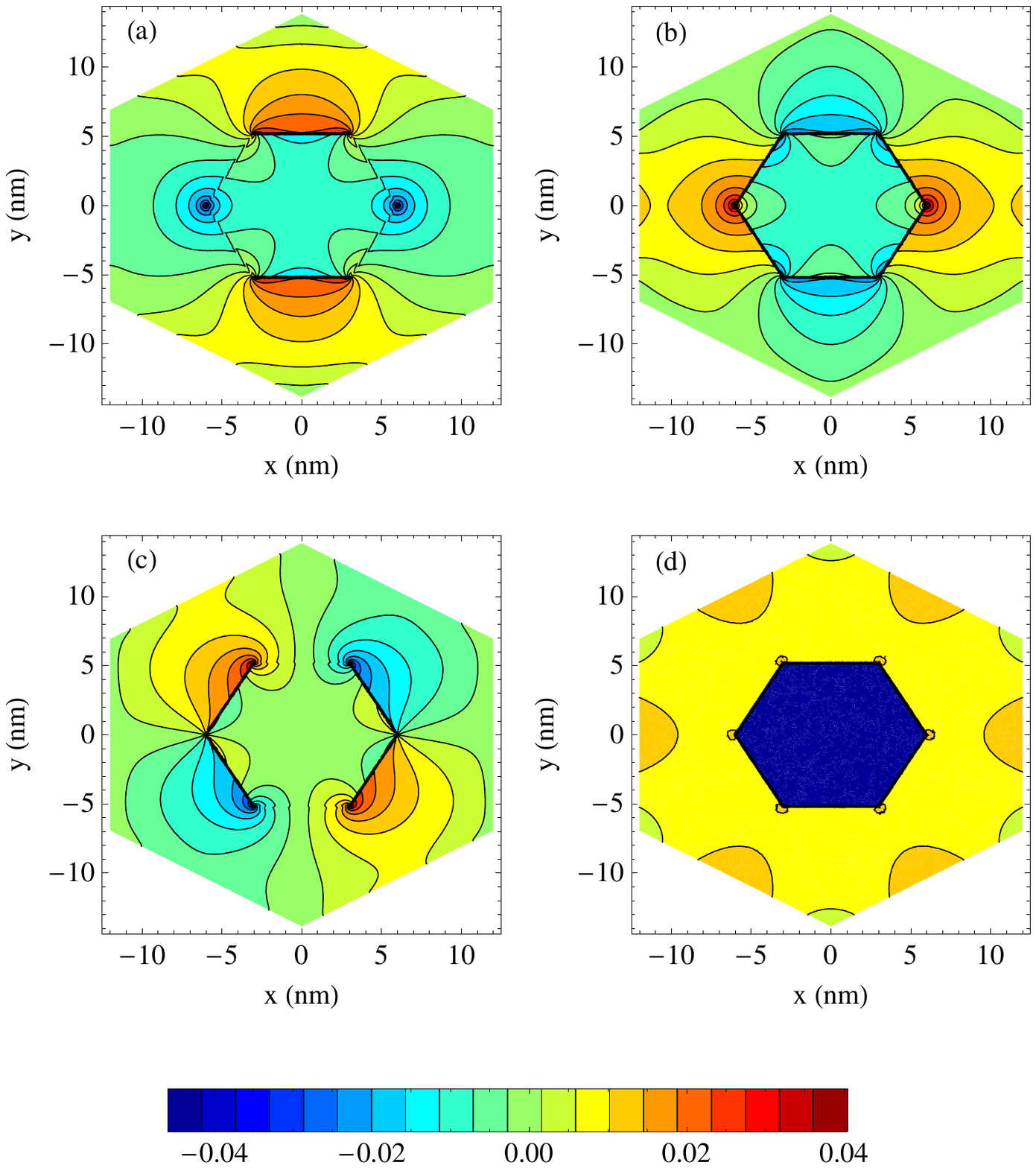}
\caption{Elastic strain distributions in a single-shell nanowire. The core with radius $r_1= 6.0$ nm is made of GaAs and the shell with radius $r_2 = 13.9$ nm is made of GaP. The plots show cross-section distributions of   (a) $\epsilon_{xx}$,  (b) $\epsilon_{yy}$, (c) $\epsilon_{x y}$, and (d) $\tr \boldsymbol \epsilon$.}
\label{fig:fig2}
\end{figure}

\begin{figure}[ht]
\includegraphics[bb=0 200 473 748,width=8.5cm]{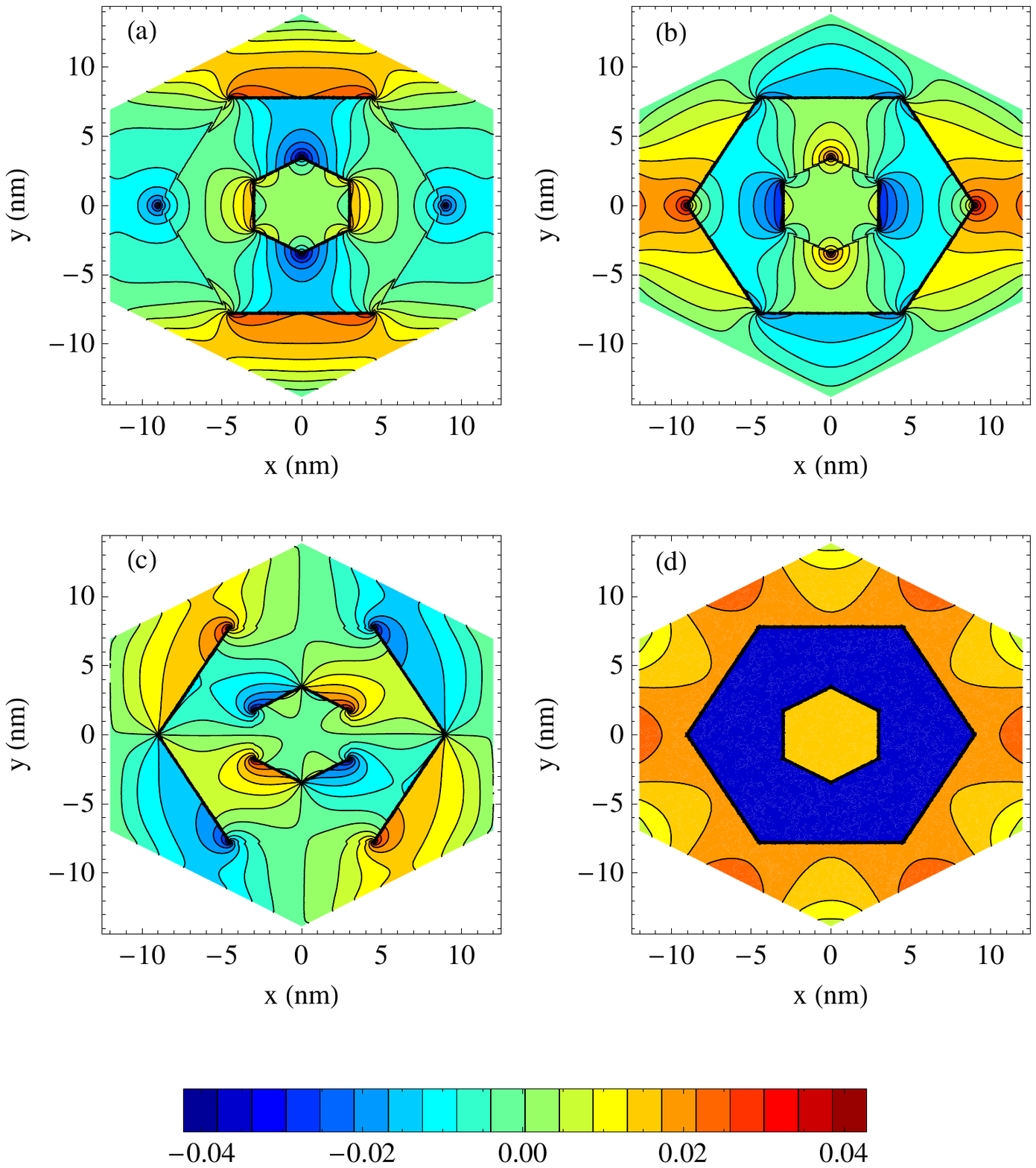}
\caption{ Elastic strain distributions in a double-shell nanowire.  The core with radius $r_1 = 3.5$ nm is made from GaP, the inner shell with radius $r_2=9.2$ nm is made from GaAs, and the outer shell with radius $r_3=13.9$ nm is made from GaP. The plots show cross-section distributions of (a) $\epsilon_{xx}$,
(b) $\epsilon_{xx}$, (c) $\epsilon_{x y}$, and (d) $\tr \boldsymbol \epsilon$.}
\label{fig:fig3}
\end{figure}

\section{Numerical results}
\label{sec:Results}
First, we consider a nanowire with a single $\text{GaP}$ shell and a
$\text{GaAs}$ core with the same geometry as in the experiment
reported in Ref.~\onlinecite{SKLPSTS05}.  We choose radii of the inner
core and shell to be $r_1=6.0$ nm and $r_2=13.9$ nm (see \refFig{fig:fig1} for
the definitions of $r_1$ and $r_2$).  We depict the strain components
$\epsilon_{xx}$ (a), $\epsilon_{yy}$ (b), and $\epsilon_{x y}$ (c), and the trace $\tr \boldsymbol \epsilon$ (d)
in \refFig{fig:fig2} (the strain component $\epsilon_{zz}$ is
omitted).  
Figure~\ref{fig:fig2}~(a) shows that near a horizontal interface the
exterior material is expanded. Figure~\ref{fig:fig2}~(b)  shows that the
exterior material is largely expanded in the regions 
near the left and right corners of the core of the hexagonal shape. These behaviours are consistent with the
difference in the lattice constants.  Figures~\ref{fig:fig2}~(a) and (b)  also show that the two strain distribution patterns are symmetric with respect to the x- and the
y-axis. The strain component $\epsilon_{zz}$, which is not shown here, exhibits a simple step-like distribution profile with a constant axial contraction in the core
and an axial elongation in the shell, again consistent with the lattice
mismatch. The shear
strain $\epsilon_{x y}$ in \refFig{fig:fig2}~(c), which is important in, e.g., the
piezo-electric interaction, displays a pattern as complicated as the
other individual strain components, $\epsilon_{xx}$ and $\epsilon_{yy}$.  The distribution 
pattern of $\epsilon_{x y}$ is, however, antisymmetric with respect to the x-axis and the
y-axis. The measure of volume
deformation, the dilatation $\delta V/V = \tr \boldsymbol{\epsilon}$,
is depicted in \refFig{fig:fig2}~(d). In general, the shell is expanded
particularly near the boundary in the middle of the edges. This
expansion extends typically all the way into the corners of the core.
The interior of the core is seen in \refFig{fig:fig2}~(d) to be contracted almost constantly.

We now turn to the case of multi-shell nanowires.  For simplicity, we
vary the setup only slightly by considering a core with {\it two}
shells, a double-shell nanowire.  We choose the material of the inner
shell to be made of GaAs and the core and outer shell are chosen to be
made from GaP. Here, in consistence with the previous geometry, we
choose the outer shell to be parallel with the inner core, see
\refFig{fig:fig1}~(c).  Still, we mention that fully parallel
multi-shells are also of interest \cite{MMF06}. The radii of the core,
inner shell, and outer shell are taken to be $r_1=3.5$ nm, $r_2=9.2$
nm and $r_3=13.9$ nm. Figure~\ref{fig:fig3} shows the strain
components $\epsilon_{xx}$ (a), $\epsilon_{yy}$ (b), and $\epsilon_{x
  y}$ (c), and the trace $\tr \boldsymbol \epsilon$ (d) of the
double-shell nanowire. (Here, again, the strain component
$\epsilon_{zz}$ is omitted, since it shows a simple step-like
distribution profile.)\ Since the lattice constant of the inner shell
is larger than its surroundings, the inner shell is compressed whereas
the core and outer shell regions are expanded (see plots for the
strain components $\epsilon_{xx}$, $\epsilon_{yy}$, and the trace $\tr
\epsilon$ in \refFig{fig:fig3}). In particular, the exterior
shell near the interface undergoes considerable stretching as in the
single-shell case. Also the same as in the single-shell nanowire case,
the strain patterns of the components $\epsilon_{xx}$ and
$\epsilon_{yy}$ are symmetric with respect to the x- and the y-axis,
and the pattern of the trace $\tr \epsilon$ is hexagonal symmetric.
Furthermore, the inner shell plays almost the same role for the
tensile strains as the inner core in the single-shell case.  The shear
strain $\epsilon_{x y}$ in \refFig{fig:fig3}~(c) shows similar
symmetric characteristics as in the single-shell nanowire, i.e., the
distribution pattern of the shear strain $\epsilon_{x y}$ is
antisymmetric with respect to the x- and the y-axis, and exhibits
similar peak-valley configurations in both inner shell and outer shell
as in the shell region of the single-shell nanowire.

\section{Summary and conclusions}
\label{sec:Summary}
We present a theoretical study of the strain field in
lattice-mismatched core-shell nanowires with single and double shells.
We derive a functional for the elastic energy in a nanowire.  The
deformation field is found by minimization of the energy functional
using the finite element method. For the single-shell wire, the core
is made of GaAs and the shell of GaP. For the double-shell wire, the
inner shell is made of GaAs, whereas the core and outer shell are made
of GaP.  We find a large volumetric strain in various regions of a
wire. A large compression appears in the core region of a single-shell
nanowire or in the inner shell region of a double-shell nanowire.  The
large volumetric strain will influence carriers via the deformation
potential.  Our numerics also shows great variations in the individual
components of the strain compared to the volumetric strain. This could
be of particular importance for the hole confinement in p-type wires.
We note that the theory presented in this work can be extended to
incorporate the finite strain components $\epsilon_{x z}$ and
$\epsilon_{y z}$ which have been neglected in the plane strain
approximation.  We leave the details of such a more general theory to
forthcoming publications.

The authors thank F. Boxberg for valuable discussions.


\begin{thebibliography}{10}

\bibitem{ZPCWL05} G. Zheng, F. Patolsky, Y. Cui, W.U. Wang, and C.M. Lieber, {\it Nature Biotechnology \bf 23}, 1294  (2005).

\bibitem{WLCWHLL04} Q. Wan, Q.H. Li, Y. J. Chen, T.H. Wang, X.L. He, J.P. Li, C.L. Lin, {\it Appl. Phys. Lett. \bf 84}, 3654 (2004).

\bibitem{LGJSY05} M. Law, L.E. Greene, J.C. Johnson, R. Saykally, and P. Yang, {\it Nature Materials \bf 4}, 455 (2005).


\bibitem{ZWM07} Y. Zhang, L.-W. Wang, and A. Mascarenhas, {\it Nano Letters \bf 7}, 1264 (2007). 

\bibitem{SDWA07} J. Schrier, D. O. Demchenko, L.-W. Wang and A. P. Alivisatos, {\it Nano Letters \bf 7}, 2377 (2007).


\bibitem{WCZZYPM08} K. Wang, J. Chen, W. Zhou, Y. Zhang, Y. Yan, J. Pern, and A. Mascarenhas, {\it Adv. Materials \bf 20}, 3248 (2008).


\bibitem{HDL05} Y. Huang, X. Duan, and C. M. Lieber, {\it Small \bf 1}, 142 (2005). 


\bibitem{CDXWWYYWG02} J. Chen, S.Z. Deng, N.S. Xu, S. Wang, X. Wen, S. Yang, C. Yang, J. Wang, and W. Ge, {\it Appl. Phys. Lett. \bf 80}, 3620 (2002).

\bibitem{GHFY06} J. Goldberger, A. Hochbaum, R. Fan, and P.D. Yang, {\it Nanoletters \bf 6}, 973 (2006). 

\bibitem{HDCLKL01} Y. Huang, X. Duan, Y. Cui, L.J. Lauhon, K.-H. Kim, and C.M. Lieber, {\it Science \bf 294}, 1313 (2001).

\bibitem{BOSPTMDWS02} M.T. Bj\"{o}rk, B.J. Ohlsson, T. Sass, A.I. Persson, C. Thelander, M.H. Magnusson, K. Deppert, L.R. Wallenberg, and L. Samuelson, {\it Appl. Phys. Lett. \bf 80}, 1058 (2002).

\bibitem{LGDWL02} L.J. Lauhon, M.S. Gudiksen, D. Wang, and C.M. Lieber, {\it Nature \bf 420}, 57 (2002).

\bibitem{LXTWL05}  W. Lu, J. Xiang, B.P. Timko, Y. Wu, C.M. Lieber, {\it PNAS \bf 102}, 10046 (2005).
\bibitem{SKLPSTS05} N. Sk\"old, L.S. Karlsson, M.W. Larsson, M.-E. Pistol, W. Seifert, J. Tr\"adg\aa rd, and L. Samulson, {\it Nano Letters \bf 5}, 1943 (2005).

\bibitem{MMF06} P. Mohan, J. Mothisa and T. Fukui, {\it Appl. Phys. Lett. \bf 88}, 133105 (2006).


\bibitem{G06} F. Glas, {\it Phys. Rev. \bf B 74}, 121302(R) (2006).

\bibitem{GSB94} M. Grundman, O. Stier and D. Bimberg, {\it Phys. Rev. \bf B 50 }, 14187 (1994).
\bibitem{GSB95} M. Grundman, O. Stier and D. Bimberg, {\it Phys. Rev. \bf B 52 }, 11969 (1995).



\bibitem{Cl03} A.N. Cleland {\it Foundations of nanomechanics} (Springer, Berlin) 2003.

\bibitem{CT93} L.D.Caro and L. Tapfer, {\it Phys. Rev. \bf B 48}, 2298 (1993).



\bibitem{PdC06} M. Povolotskyi and A. Di Carlo, {\it J. Appl. Phys. \bf 100}, 063514 (2006).

\bibitem{ZT94} O.C. Zienkiewicz and R.L. Taylor {\it The finite element method} (McGraw-Hill, Maidenhead) 1994.


\bibitem{L94}  A.E.H. Love  
                {\it A treatise on the mathematicl theory of elasticity}
                (New York: Dover) 1994.

\bibitem{AM76} N.W. Ashcroft and N.D. Mermin {\it Solid State Physics} (Saunders College, Philadelphia) 1976.





\end{thebibliography}
\end{document}